\documentclass[manuscript,nonacm]{acmart}
\usepackage[boxed,noline,noend]{algorithm2e}
\begin{document}

\setcopyright{acmcopyright}
\copyrightyear{2022}
\acmYear{2022}

\acmConference[FAccT '22]{FAccT '22}{June 21--24, 2022}{Seoul, South Korea}

\title[From Data Leverage to Data Co-Ops: An Institutional Model]{From Data Leverage to Data Co-Ops: An Institutional Model for User Control over Information Access}

\author{Caleb Malchik}
\email{caleb.malchik@yale.edu}
\affiliation{%
        \institution{Yale University}
        \city{New Haven}
        \state{CT}
        \country{USA}}

\author{Joan Feigenbaum}
\email{joan.feigenbaum@yale.edu}
\affiliation{%
        \institution{Yale University}
        \city{New Haven}
        \state{CT}
        \country{USA}}

\begin{abstract}
Internet companies derive value from users by recording and influencing their behavior.
Users can pressure companies to refrain from certain invasive and manipulative practices by selectively withdrawing their attention, an exercise of \textit{data leverage} as formulated by Vincent {\it et al}.
Ligett and Nissim's proposal for an institution representing the interests of users, the \textit{data co-op}, offers a means of coordinating this action.
We present one possible instantiation of the data co-op, including the Platform for Untrusted Resource Evaluation (PURE), a system for assigning labels provided by untrusted and semi-trusted parties to Internet resources.  We also describe PURESearch, a client program that re-ranks search results according to labels provided by data co-ops and other sources.
\end{abstract}



\maketitle

\section{Introduction}\label{sec-introduction}

\subsection{Data Co-Ops}\label{subsec-dataco-ops}

The Data Co-ops initiative of Ligett and Nissim~\cite{Ligett20} is a
wide-ranging, multi-institutional effort aimed at exploring the
relationships between Web users and digital platforms.  During the past
decade, a number of smaller-scale initiatives have focused on issues like
collecting higher-quality data, (re-)gaining individual control over
personal data, (re-)establishing trust in the online information environment,
generating rich data for research purposes, \textit{etc.}  Each of them
addressed one or more specific shortcomings of the status
quo, \textit{e.g.}, privacy risks, unrealized potential of data, individuals’
receiving limited value in exchange for their behavioral data, and lack of
opportunities for public governance of digital creation.  By contrast, the
Data Co-ops initiative seeks more broadly to address, in a technologically
comprehensive fashion, users' privacy concerns and the potential to tap
as-yet-unrealized value in users’ behavioral data.

In our formulation, a data co-op is a membership organization that
provides client software, a means of resource discovery, and technical
support.  There are many possible revenue models for such an organization,
the simplest of which is to have the members pay dues. A data co-op
could provide means for users to communicate or publish content, such
as digital identities and hosting resources. Thus, in the short run, the
client software provided by data co-ops as we conceive them would advance
Ligett and Nissim’s goals of increased user  privacy and more widely
distributed value creation. Longer term, a data co-op could provide the
means to participate fully in a new network, based on an improved web
architecture, as the network grows.

As a first step, we have created a tool that allows co-op members to
shape their own processes of resource discovery on the existing Web.

\subsection{Implicit User-Provider Negotiations}
\label{subsec-implicit-negotiations}

When a user interacts with an Internet service under typical market
conditions, the user and the provider both want to maximize the value
they derive from the interaction while minimizing costs.  The interests
of the two parties are aligned in some aspects, such the desire to meet
the needs of the user so that she continues to use the service.
In other aspects, their interests conflict; for example, in an ad-supported
service, the provider wants to maximize the data collected
about the user, the time and attention spent on ads, and the capacity
to influence user behavior, but the typical user wants to minimize these things.

The adversarial component of this relationship leads to an implicit
negotiation: Users want to fulfill their needs while offering the
provider only enough value to sustain the service.  The provider wants
to maximize value extraction, while offering the user only enough value
to motivate her to continue to use the service.  The strong market
positions of many of today's providers, together with the relative
indifference and atomization of users, causes these interactions to tend
toward the latter extreme.

This negotiation resembles that between buyers and sellers, and
between employees and employers (buyers and sellers of labor).  In the
same way that unions and consumer cooperatives can shift negotiations
in favor of employees and consumers, data co-ops could shift these
implicit negotiations over data, attention, and services
in favor of users, leading to services that are less invasive,
manipulative, and addictive, as well as more useful and reliable.

Although a data co-op could take many different approaches to achieving this
goal, our current focus is on technical interventions. In particular, we
consider the use of labels to direct traffic towards well behaved
pages, services, and other network resources and away from poorly behaved ones.

\subsection{Types of Intervention and Prospects for User Control}
\label{subsec-type-of-interventions}

Efforts to address the harms of mass data collection have come in
many forms: technical interventions such as Tor~\cite{dingeldine04tor} and
Pretty Good Privacy~\cite{garfinkel1995pgp}, legal
interventions such as the European Union's General Data Protection Regulation,
and interventions by advocacy organizations such as the
Electronic Frontier Foundation's ``report cards'' and the Free Software
Foundation's promotion of free software.  These
categories are not strictly disjoint: Any non-technical intervention
requires technical efforts to implement, and any technical intervention
requires non-technical efforts to support its use.

In the context of a data co-op, the most interesting approaches
involve negotiating with service providers (implicitly or explicitly)
to offer services in a certain way, and facilitating the use of certain
client software by co-op members.

It is helpful to examine the free-software norm in view of the dynamic
described in the previous section.  Free software offers a protective mechanism
against anything that the user would not want a program to do,
because users can modify the program or pay someone to modify it for
them. However, this requires nearly boundless resources (while users do not even
have time to read the terms of use for most services), and modifying client
software can introduce incompatibilities with the server.

The shortcomings of the free-software norm are illustrated by the
evolution of the Web. As the Web grew more complex and more dominated
by huge companies, the complexity of browsers reached a point at which even a
free browser like Firefox is beyond the control of users, because it
requires a large team of developers to make changes without breaking
things.  Even if this barrier were overcome, many websites require JavaScript,
which is often obfuscated.  Even if you did examine the JavaScript code
on every website that you visit and modify it to prevent needless data
collection or manipulative user interface elements, these changes could break
the website's functionality.

The Web example also demonstrates that effective control over what
data are sent to the server requires regulation of the entire client
program, not just the network protocol.  HTTP was designed for simple
hypertext, which lacks many of the data-collection capabilities that
modern websites employ, but HTTP can still transport tracking scripts
that are run by a browser engine and report behavioral data back to the
server (also via HTTP). Designing the protocol for simple hypertext did
not prevent these innovations.

For a group of users to maintain control over the client program,
the client code that is specific to a service provider should be
minimized.  This could be accomplished by keeping to declarative-markup
languages, where it is easy to understand which user actions cause data
to be sent to the server and what data are sent.

While users still depend on client software that accommodates data
collection and other harms, labeling can allow users to preferentially
use less harmful sites.  This approach also provides a path to gaining
control over the client software by filtering out resources that require
software developed by unaccountable third parties.

\subsection*{Paper outline}

In Section~\ref{sec-related-work}, we situate this work
within the FAccT research agenda and present related work.
In Section~\ref{sec-re-ranking}, we describe the Platform for Untrusted Resource Evaluation (PURE), a framework for assigning labels to online resources, and PURESearch, a tool that re-ranks search results according to PURE labels provided by the user
and semi-trusted third parties. In Section~\ref{sec-adoption}, we argue that
a data co-op's offering PURE labels and related software to its members
would represent a good value proposition and thus that widespread adoption of
PURE or something similar is plausible.
Finally, Section~\ref{sec-future} concludes the paper and discusses future directions.

\section{Related Work}\label{sec-related-work}

We begin this section by situating data co-ops and PURE
within the broader FAccT agenda, paying particular attention to
the work of Vincent~{\it et al.} on data leverage~\cite{Vincent21}.  We then
briefly discuss three other areas of research that our work draws on.

\subsection{Data Co-ops and the FAccT Agenda}\label{subsec-situate}

Sociotechnical systems play a central role in the FAccT community's goal of
achieving fairness, accountability, and transparency in the online world.
Roughly speaking, sociotechnical systems are technologies that support and
influence people's daily lives by taking a multidisciplinary approach.
PURE is a sociotechnical system that seeks
to alter the balance of power described in
Section~\ref{subsec-implicit-negotiations} so as to better position users in
their implicit negotiations with providers. As noted in
Section~\ref{subsec-type-of-interventions}, the sociotechnical approach is the
natural one to take when trying to increase users' control over their
behavioral data, because social mechanisms intended to increase control will
need technological instantiation, and technical mechanisms will not be
adopted unless they are enforced legally, incentivized financially, or
supported by social trends or norms.

The potential effectiveness of data co-ops and our particular approach
to them can be understood and evaluated using the data-leverage framework of
Vincent~{\it et al.}~\cite{Vincent21}, the goal of which is to ``highlight
new opportunities to change technology company behavior related to privacy,
economic inequality, content moderation, and other areas of social concern.''
Three levers that are available to users are identified in~\cite{Vincent21}:
{\it data strikes}, in which users withhold or delete data to reduce the
efficacy of an organization's data-dependent technologies; {\it data
poisoning}, in which users insert inaccurate or harmful data into
an organization's data-dependent technology; and
{\it conscious data contribution} (CDC), in which
users give their data to organizations they support.
Data co-ops can provide their members with data leverage.  In particular,
they can manage labels associated with websites, thus
effectively subjecting certain sites to boycott; in this way, co-ops can
coordinate data strikes, data poisoning, and CDC by their members.  Labels also
allow this point of leverage to be used more fluidly and continuously
by subtly steering users away from misbehaving sites and towards more favored
ones.

Pentland and Hardjono~\cite{Pentland20202}, in work that is independent of
Ligett and Nissim's, have also put forth the idea that
data co-ops can rebalance the world's data economy.

\subsection{Labels}\label{subsec-labels}

Labels are commonly applied to static content in the context of parental
controls, but they can incorporate arbitrary information and be applied
to many types of online resources.  To protect user data and improve
Web usability, labels could encode information about privacy policies,
popups, resource usage, compatibility with privacy-protecting client
software, and more.

Previous labeling systems such as the
Platform for Internet Content Selection (PICS) require labeling services
to be trusted:  "When publishers are unwilling to participate, or can't
be trusted to participate honestly, independent organizations can provide
third-party labels"~\cite{Resnick96}.  This creates a barrier to
participation as a source of labels and limits the range of labels that
can safely be used.

PURE
improves on this by emphasizing labels that can be verified
by end users and leveraging user inputs to assign a reputation
to each label source.  Under this regime, publishers who self-label
their content can be held accountable by the ecosystem of users and
label sources.

\subsection{Data as Labor}\label{subsec-labor-capital}

Many works, notably those of Posner and Weyl~\cite{Posner18}
and Arrieta-Ibarra~{\it et al.}~\cite{Arrieta-Ibarra18}, promote the
concept of ``data as labor.'' Currently, corporations that monetize users'
data view those data as their property that {\it they create} by providing
services to people who use them willingly. Adherents to the data-as-labor
school of thought view data as valuable products that {\it users create} and
that corporations profit from. Their view leads naturally to the goal expressed
in Section~\ref{sec-introduction} and in \cite{Vincent21}: Rebalance the
relationship between the users who create data and the corporations that
profit from them so that users have more knowledge about which data are
collected, more control over how data are monetized, and more ability to
reap the rewards of data-dependent commerce. Data co-ops may thus be viewed
as analogous to labor unions, which serve precisely this function in
negotiations between workers and their employers. This analogy was drawn
explicitly by Pentland and Hardjono~\cite{Pentland20202}; similarly,
Posner and Weyl suggest the formation ``data unions.''

\subsection{Recommender Systems}\label{subsec-recommender-systems}

Recommender systems~\cite{Jannach10} are widely deployed in social media and
entertainment platforms to predict which content will engage the user
most effectively.  PURE
can be seen as an unusual type of
recommender system that maximizes the user's control over what is
"recommended." Typical recommendation systems employ either or both of two
broad approaches: content-based filtering and collaborative filtering.
Both are similar in some ways to the approach taken with PURE,
but neither applies in a conventional way.

Content-based filtering uses a set of predefined labels for each item,
such as a database of song attributes used to recommend music.  Typically
the user gives feedback on each item via binary or unidimensional ratings,
and an algorithm builds a model of which labels the user prefers.
Our approach is almost the opposite: Instead of predicting which
labels a user likes, we predict which labels an item has.

Collaborative filtering recommends content to a user based on similar
users' responses to the same content.  Determining similarity between
users is almost the same as determining agreement between label sources,
and complex algorithms for collaborative filtering could be used in place
of our simple algorithm.  The main difference is that we use agreement
between sources to estimate the values of labels, not the overall rating for
a resource.

Our approach is also distinguished by doing all processing of labels
and ratings on the client side, resulting in less transmission of
user data.

\section{Re-ranking Search Results Based on Third-party Labels}
\label{sec-re-ranking}

As an initial technical intervention, we have developed PURESearch, a tool that re-ranks search results according to labels provided by the user and semi-trusted third parties.  This allows data co-ops and their members to direct usage towards certain web resources and away from others, improving the quality of the results and pressuring publishers to follow certain practices.

In the current (simplest) form, PURE labels are uninterpreted strings that can be bound to URLs with an affirmative or negative assertion, depending on whether the label applies or does not apply.  A client program is configured to promote resources with certain labels and to demote others, according to the user's preferences (or a standard configuration received from a data co-op).  The client fetches search results from an upstream search engine, and re-ranks them according to label preferences.

Potentially useful labels include \textit{haspopup}, \textit{hasfixednavbar} and \textit{hascookiebanner}, which apply respectively if the page has a popup, a fixed navigation bar, or a cookie banner.  Users may want to filter or demote pages with such labels for the sake of usability.  Cookie banners also have implications for data collection, as the user may be required to agree to certain terms in order to remove the obstruction from the page.  Mass filtering of pages with cookie banners that do not require cookies for their functionality would pressure publishers to remove such banners and only use cookies when necessary.

The aforementioned labels are helpful when using a standard modern web browser, but a more durable solution would be to use client software that doesn't support unwanted features.  Labels such as \textit{noscriptcompat} and \textit{lynxcompat} could apply respectively if the page is compatible with the NoScript browser addon~\cite{NoScript} or the Lynx browser~\cite{Rakitin97}.  NoScript blocks all scripts on a page, preventing many forms of data collection and reducing resource usage, but this aggressive measure often renders pages unusable.  Lynx is a text-mode browser without support for scripts, which has particular utility for the blind by ensuring compatibility with screen readers~\cite{seltzer1995maintaining}.  Users of NoScript, Lynx, or similar software could improve their browsing experience by filtering out pages that are incompatible with their chosen client software.  This improvement of the experience could induce more users to use such software, putting pressure on publishers to accommodate them.

\subsection{Search Client Overview}\label{subsec-search-client-overview}

\begin{figure}[h]
\centering
\includegraphics[width=\linewidth]{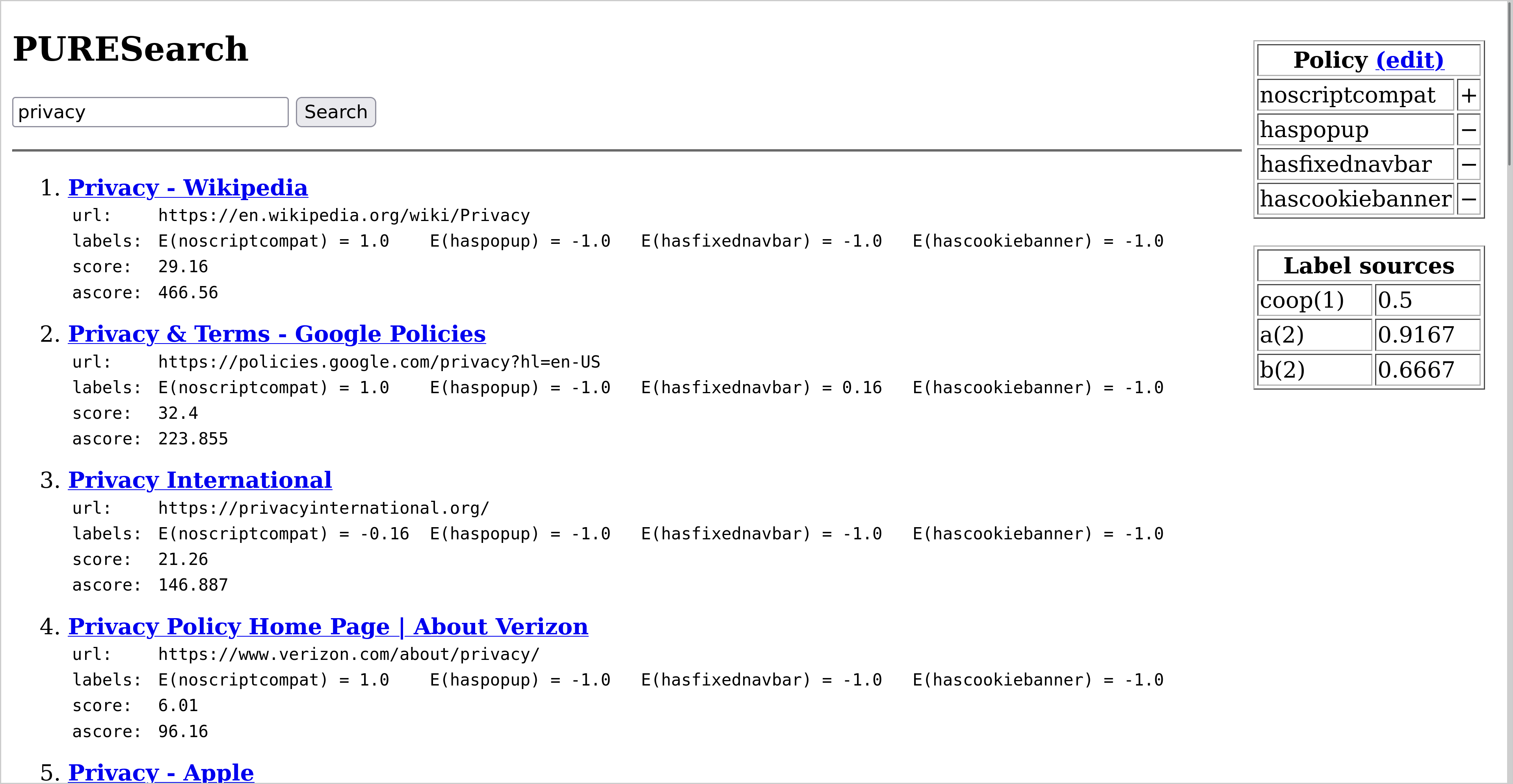}
\caption{PURESearch results for "privacy," including auxiliary information for each source.  Here, "score" refers to the upstream relevance score and "ascore" refers to the adjusted score used for re-ranking.  The tables to the right show the user's label preferences and the configured sources along with their tier numbers and reputation scores.
\medskip 
}
\Description{A web page with the heading "PURESearch" and a search bar with the text "privacy" followed by a list of search results.  To the right are two tables, one titled "Policy" followed by a link with the text "(edit)," and the other title "Label sources."}
\label{fig:search}
\end{figure}

\begin{figure}[h]
\centering
\includegraphics[width=\linewidth]{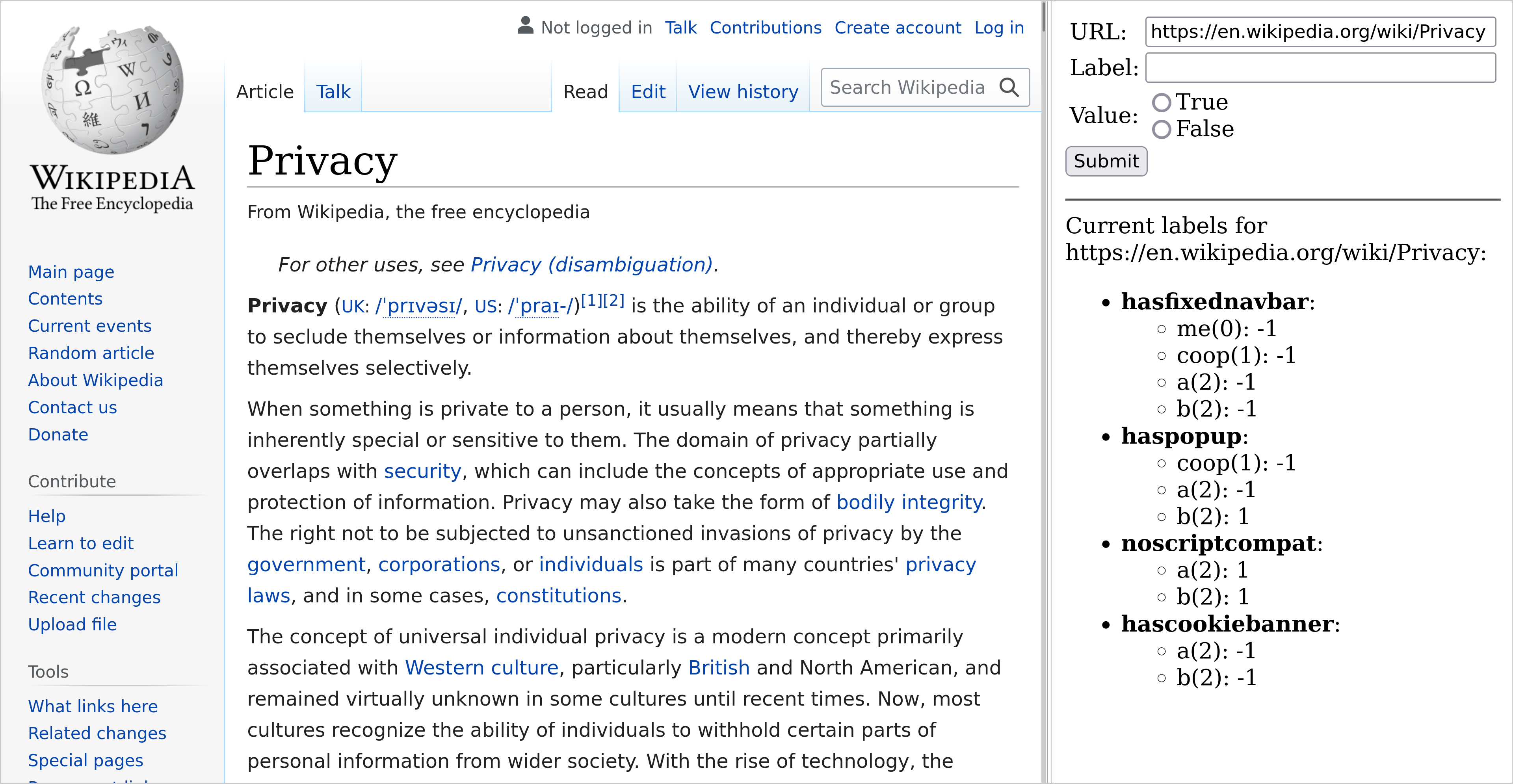}
\caption{Sidebar for the user to inspect label records for a page and enter their own.}
\Description{A web page divided into two frames, one containing the Wikipedia page for Privacy and the other containing a form with the fields "URL," "Label," and "Value," followed by a list of existing labels.}
\label{fig:sidebar}
\end{figure}

The search interface is implemented as an HTTP service that is intended to
run on the client machine.  It could be adapted to run on a remote server,
but keeping as much as possible on the client side eliminates unnecessary
transmission of user data and improves performance and reliability.

We still rely on a remote search engine (an instance of Searx~\cite{Searx}) for
results, and query strings are transmitted to the search engine from
the user's IP address.  This could be improved by hosting parts of the
search engine infrastructure on semi-trusted servers, possibly operated
by a data co-op.

Typical usage of the program proceeds as follows:

\begin{itemize}
\item The user configures which labels should be promoted in search results
  and which should be suppressed.
\item Label records are retrieved from a list of sources and processed to
  estimate the degree to which each label applies to each URL (since
  sources may not be trustworthy and label records may conflict).
\item Search results are fetched from a search engine and reordered according
  to labels and user preferences (see Figure~\ref{fig:search}.
\item When the user views a result, she may open a sidebar to inspect the
  label data for that URL and enter their own label assertions (see
  Figure~\ref{fig:sidebar}).
\item Label assertions by the user are treated as ground truth, and the
  weight given to each source when estimating labels is determined by
  the degree to which a source's labels agree with the ground truth.
  Sources may also be categorized in tiers, allowing sources in higher
  tiers to serve as ground truth for sources in lower tiers.
\end{itemize}

\subsection{Label Retrieval and Processing}\label{subsec-label-processing}

\begin{algorithm}
\SetKwFunction{Reputation}{Reputation}
\SetKwFunction{Expectation}{Expectation}
\SetCommentSty{textrm}
\SetFuncArgSty{textrm}
\SetFuncSty{textrm}
\SetArgSty{textrm}
\DontPrintSemicolon
\SetKw{Continue}{continue}
\textbf{Function} Reputation($j$)\;
\vspace{-0.1\baselineskip}\hrule\vspace{-0.9\baselineskip}\;
\tcp{Input: label source $j$}
\tcp{Output: the reputation of $j$, ranging from 0 to 1}
\tcp{$t_j$ is the tier of $j$, with 0 denoting the highest tier}
\If{$t_j = 0$}{
	\Return{1}
}
$n \leftarrow 0$\;
$d \leftarrow 0$\;
\For{$(i, k, v_{ijk})$ such that $j$ has given $v_{ijk}$, either 1 or $-1$, in evaluating item $i$ for label $k$}{
	\If{$\Expectation(i, k, t_j - 1) > 0$}{
		$x \leftarrow 1$\;
	}
	\ElseIf{$\Expectation(i, k, t_j - 1) < 0$}{
		$x \leftarrow -1$\;
	}
	\Else(\tcp*[h]{$\Expectation(i, k, t_j - 1) = 0$}){
		\Continue
	}
	\tcp{add 2 when $j$ disagrees with higher tiers}
	$n \leftarrow n + |v_{ijk} - x|$\;
	$d \leftarrow d + 1$\;
}
\If{$d = 0$}{
	\Return{0}
}
\Return{$\max(1 - \frac{n}{d}, 0)$}
\caption{Estimating the trustworthiness of a label source.  Note the mutual recursion with the Expectation function.}
\label{alg:reputation}
\end{algorithm}

\begin{algorithm}
\SetCommentSty{textrm}
\SetFuncArgSty{textrm}
\SetFuncSty{textrm}
\SetArgSty{textrm}
\DontPrintSemicolon
\textbf{Function} Expectation$(i, k, t)$\;
\vspace{-0.1\baselineskip}\hrule\vspace{-0.9\baselineskip}\;
\tcp{Input: item $i$, label $k$, and tier $t$}
\tcp{Output: the expected value of label $k$ for item $i$, ranging from -1 to 1, based on sources from tiers 0 through $t$}
\If{$t < 0$}{
	\Return{0}
}
\If{$\Expectation(i, k, t - 1) \neq 0$}{
	\Return{$\Expectation(i, k, t - 1)$}
}
$n \leftarrow 0$\;
$d \leftarrow 0$\;
\For{$(j,v_{ijk})$ such that source $j$ has tier $t$ and has given $v_{ijk}$ in evaluating item $i$ for label $k$}{
	$n \leftarrow n + \Reputation(j) \times v_{ijk}$\;
	$d \leftarrow d + \Reputation(j)$\;
}
\If{$d = 0$}{
	\Return{0}
}
\Return{$n/d$}
\caption{Estimating the applicability of a label.}
\label{alg:expectation}
\end{algorithm}

\begin{algorithm}
\SetCommentSty{textrm}
\SetFuncArgSty{textrm}
\SetFuncSty{textrm}
\SetArgSty{textrm}
\DontPrintSemicolon
\textbf{Function} Adjustment($i$)\;
\vspace{-0.1\baselineskip}\hrule\vspace{-0.9\baselineskip}\;
\tcp{Input: item $i$ from search results, in this case a URL}
\tcp{Output: factor by which to adjust relevance score for $i$}
$r \leftarrow 1$\;
\For{each label $k$ in user policy}{
	\If{$k$ is favored}{
		\tcp{$t_{max}$ is the lowest tier, denoted}
		\tcp{by the highest number}
		$q \leftarrow \Expectation(i, k, t_{max})$\;
	}
	\Else(\tcp*[h]{$k$ is disfavored}){
		$q \leftarrow -\Expectation(i, k, t_{max})$\;
	}
	\If{$q \geq 0$}{
		$r \leftarrow r(1 + q)$
	}
	\Else{
		$r \leftarrow r(1 + \frac{q}{1-q})$
	}
}
\Return{r}
\caption{Calculating the factor by which to adjust the relevance score of a search result.}
\label{alg:adjustment}
\end{algorithm}

For simplicity, a remote label source is represented by a URL, which the
client polls for the current version of that source's labels.  Each label
source is assigned a tier number, with the user having tier 0, the highest
tier, and higher numbers denoting lower, more subordinate tiers.

Label records are stored in text files with one record per line, each
record consisting of a label name, a value (1 if the label applies,
-1 if it does not), and a URL, separated by tabs.  The format is used
for remote label sources as labels are fetched as well as local storage
of labels from both remote sources and the user.

Labels are processed to establish a reputation for each label source
as detailed in Algorithm~\ref{alg:reputation}.  On the last line of
the algorithm, $\frac{n}{d}$ is double the proportion of (label, URL)
pairs for which the label source $j$ disagrees with sources in higher
tiers, so the reputation of a source is 0 if it disagrees with sources
in higher tiers more than half the time.

After fetching search results, label data that are relevant to the
results and the user's preferences are processed according to Algorithm
\ref{alg:expectation} to produce an expected value between 1 and -1
for each (label, URL) pair, with 1 representing certainty that the label
applies, -1 representing certainty that it does not, and 0 representing
absolute uncertainty.  This expected value is an average of the label
values from each source, weighted by each source's reputation.

Each upstream search result includes a relevance score which determines
the upstream order of results.  We rearrange the results by adjusting
relevance scores according to Algorithm~\ref{alg:adjustment}.  Here, $q$
represents the \textit{favorability} of the item with respect to label
$k$, so that a positive expectation of a favored label yields the same
effect as a negative expectation of a disfavored label with the same
level of certainty.  The running product is multiplied by $1 + q$
for a favorable assessment and $1 + \frac{q}{q-1}$ for an unfavorable
assessment, so that an unfavorable assessment cancels out a favorable
assessment with the same magnitude.

Note that checks for a zero value of the Expectation function in
Algorithms~\ref{alg:reputation} and \ref{alg:expectation} treat even
disagreement between sources in a given tier the same as an absence of
labels from that tier.  Furthermore, a nonzero expectation from tier $t$
supercedes any lower tiers (with numbers greater than $t$), even if it
has a small magnitude indicating uncertainty.  Note also that sources
with zero reputation are ignored regardless of tiers.  These finer points
were chosen arbitrarily or for the sake of simplicity, and could just as
well have been done differently.  Changes to the handling of reputations
and tiers could make more sense {\it a priori} in certain contexts,
or simply yield better results.

A proof-of-concept implementation and self-guided
demonstration of PURESearch can be downloaded at
\linebreak 
\texttt{http://cs.yale.edu/homes/cmalchik/puresearch-0.1.tar.gz}.

\subsection{Possible Extensions}\label{subsec-possible-extensions}

The version of PURESearch presented here is meant to give a clear
illustration of the purpose of PURE labels and what it means for label
sources and rated items to be "untrusted."  Simplicity is necessary in
general to maximize the ability of users or trusted technical experts
to understand and control the software they run.  However, extensions to
the current version of PURESearch could certainly improve its usefulness.

Different label processing algorithms, possibly adapted from prior
work in recommender systems, could be used for different subsets of
label records.  For instance, the Influence Limiter of Resnick and
Sami~\cite{resnick2007influence} limits the capacity for manipulation by a
malicious rater able to create a bounded number of sybils.  This would be
well suited to a special lowest tier of label sources which are imported
without manual vetting, possibly including individuals who publish their
label records to a public registry.

Labels could have a continuum of values rather than binary 1 or -1, which
could signify uncertainty on the part of the label source or ambiguity
in the applicability of the label.  Values could also include symbols
that specify the semantics of the assertion, describing how the label
relates to the content rather than just the degree to which it applies.
These extensions raise the cost of verification because complex values
may take more time to determine than a simple "yes" or "no."

Labels could also refer to properties that can not be verified by an end
user looking at the content, such as authorship or copyright status,
or aggregate ratings of privacy practices.  This would require other
accountability mechanisms to make the labels trustworthy.  For example,
lying about the copyright status or origin of a work could incur legal
liability.  Unverifiable labels could also be retrieved from trusted
sources such as a data co-op with strong internal accountability
mechanisms.

Labels could also be extended to apply to classes of URLs as defined by
regular expressions, prefixes, or some other method.  A na\"ive approach
to this would disrupt the ability to add or correct labels based on
interaction with a single page; a more nuanced approach could treat a
label record about classes of URLs as a series of ordinary label records.

Finally, label sources could be enabled to make higher order assertions
about other label sources, enabling delegation, detraction, and
third-party vetting.  This would warrant careful limits on the types of
assertions allowed, to prevent label records from becoming too difficult
to understand.

More extensive changes to PURE are discussed in Section~\ref{sec-future}.

\section{Value Proposition and Plausibility of Adoption}\label{sec-adoption}

In this section, we explain why we believe that data co-ops and PURE constitute a good value proposition for users and a plausible path to a fairer and more productive balance of power in the data economy.

\subsection*{Ease of use and low overhead}
Although many people care about lack of privacy and inability to exploit valuable data that they themselves created, few people care {\it enough} about this pervasive unfairness to be willing to put thought and effort into combatting it on an ongoing basis.  As the description of ``typical usage'' in Section~\ref{subsec-search-client-overview} makes clear, no ongoing thought and effort is required for a co-op member to use PURE. Users need not configure their own search interfaces; they can use the default configuration that is provided and updated by the co-op’s technical staff.  Similarly, users {\it may} enter their own label assertions, but they are not required to do so.

We expect there to be some users who regularly enter label assertions and, more generally, act consciously to advance the mission of the co-op. A small number of users with technical skills and strong commitment to the cause may join the co-op’s technical staff as paid employees.  One sees the same range of engagement in labor unions: Most people who work in unionized industries simply join the union, pay their dues, and reap the benefits of collective bargaining; some are more active in the union’s negotiations with the employer or in its political or social activities; and a few seek paid employment in official union-leadership positions.

\subsection*{An idea whose time has come} 
As explained in Section~\ref{subsec-situate}, technical mechanisms like PURE are rarely adopted unless they are enforced legally, incentivized financially, or supported by social trends or norms. For several years, our society has been trending toward resentment of Big Tech and hunger for protection against Big Tech’s predations.  Data co-ops can provide protection as well as leverage for users who wish to channel their resentment constructively. Co-ops are also well situated to support socially beneficial norms of online behavior and to provide forums for discussion and evolution of such norms.

\subsection*{Broad applicability} 
The PURE approach is optimized for flexibility. Labels can apply to any networked resource with a name, and the labels themselves can refer not only to privacy or control of data but to any property that users care about.  For example, websites can be labeled according to how well they deal with various forms of harmful content over which there has been recent public concern: false or misleading information that may have contributed to recent political turmoil in many western countries; addictive and manipulative applications and services, which may negatively impact the mental well being of users, especially teens; and misinformation about COVID19 vaccines. 

Labels with varying semantic properties could help address each of these issues. A diverse range of data co-ops could produce or curate labels for different domains and purposes. Labels may also be provided by individuals and other types of institutions; of particular interest might be labels that are computed by technologically sophisticated, large-scale ``Internet observatories’’ that measure and analyze phenomena that cannot be observed by a single user or even a typical co-op, which may be relatively homogeneous geographically or demographically. By using a semantically diverse and powerful range of labels, a co-op could, for example, deploy a singular browser extension that marks certain pieces of content as false or misleading. The more ambitious goal of comprehensively optimizing the Internet experience on behalf of co-op members may also be within reach. 

In summary, the PURE labeling framework is transparent with respect to the semantics of each label and the objects that labels may apply to.  It can serve a very broad range of uses\ --\ not only data co-ops as we have conceived them.

\subsection*{Protocol independence and support for alternative protocols}

PURE labels are separate from and independent of the resources they
refer to, so the issue of compatibility between a label record and its
object does not arise as long as the object has a name.  This shields
PURE from the burden of maintaining compatibility with the Web as it
rapidly evolves.  It also makes PURE well suited to alternative protocols
such as Gemini~\cite{Gemini} and Gopher~\cite{Kaiser, Anderson09}, which
have developed niche followings as a result of discontent with the modern
Web and the impossibility of maintaining a modern Web browser without
significant capital investment.

A data co-op could act as an incubator for an alternative protocol by
curating PURE labels and providing client software for the protocol.
The concept of a client-side resource discovery tool, illustrated by our
PURESearch program, suggests software that displays resources using an
alternative protocol alongside the more familiar and numerous resources
on the existing Web.

\section{Conclusions and Future Directions}\label{sec-future}

In this work, we have presented the PURE labeling framework as a powerful
and flexible way to promote online privacy and usability, and to increase
the share of value that users receive from the data that they create.
We have also implemented PURESearch, a client-side search tool based
on this framework. To meet the need for an institutional and social
setting in which to deploy PURE technologies, we proposed data co-ops
as they were conceived by Ligett and Nissim~\cite{Ligett20}, and argued
that they can play a natural and constructive role in the FAccT agenda.

While the current iteration of PURESearch is designed to improve usage
of the existing Web, PURE establishes a basis for much more radical
changes to Internet usage.  The techniques for supporting alternative
protocols discussed in Section~\ref{sec-adoption} could be applied to
a new protocol designed specifically for use with PURE and data co-ops.

We believe such a protocol would be well served by a proposed
future Internet architecture called Named Data Networking (NDN)
\cite{zhang2014named}.  NDN offers a way to make the Internet more
useful and reliable, while making it more difficult to collect data
access patterns on a mass scale.  This is due to in-network caching and
the lack of source information in data request packets.  A data co-op
operating in a geographic locality would be well positioned to operate a
set of NDN nodes, eliminating the bulk of external requests for popular
content while saving bandwidth and enhancing the user experience.

Immutable named data objects (NDOs), a core component of NDN, also form an
attractive basis for labeling.  URLs are a rough identifier for content
because the content accessed at a URL can change depending on when it
is accessed and the IP address or user agent of the requester.  In fact,
there is nothing preventing a web server from serving different content
for each request.  This makes URLs a shaky basis for labels that are meant
to refer to a particular piece of content.  NDOs do not have this problem
because they consist of immutable data objects cryptographically bound to
a canonical name and the publisher's public key.  Resources organized as
NDOs would thus be much better suited to labeling, and dishonest labelers
would no longer have plausible deniability when issuing faulty labels.

In addition to a new data transport protocol, we see potential utility
in a new semantic hypertext language designed with PURE and data co-ops
in mind, particularly with respect to the goal of user control over
client software discussed in Section~\ref{subsec-type-of-interventions}:
The behavior of an HTML document running JavaScript is impossible to
determine in a bounded amount of time; bounds on the resource usage of
scripts can allow for some reasoning about the properties of a page,
but the details remain fundamentally opaque.  HTML and JavaScript
could be replaced by a purely declarative markup language, giving the
client program full insight into what a page is doing.  Gemini shares
this goal but lacks functionalities that users of the modern Web have
come to expect.  Such functionalities, which are traditionally left
to scripts, could be replaced by declarative markup with well-defined
behavior, giving the user full control over the behavior of the client.
Extensions to standard markup could be evaluated and labeled by data
co-ops or third parties, with assurances that a markup extension includes
no unnecessary data collection or other malicious features.

\bibliographystyle{ACM-Reference-Format}
\bibliography{db}

\end{document}